\documentstyle[aps,epsfig,twocolumn]{revtex}

\begin{document}

\draft

\title{ENTROPIC NONEXTENSIVITY:  A POSSIBLE MEASURE OF COMPLEXITY}
\author{Constantino Tsallis}
\address{Centro Brasileiro de Pesquisas Fisicas, Xavier Sigaud
150, 22290-180 Rio de Janeiro-RJ, Brazil\\
Department of Physics, University of North Texas, P.O. Box 311427, 
Denton, Texas 76203, USA \\
Santa Fe Institute, 1399 Hyde Park Road, Santa Fe, New Mexico 87501, 
USA\\
tsallis@cbpf.br\\
$\;\;\;\;\;\;\;\;\;\;\;\;\;\;\;\;\;\;\;\;\;\;\;\;\;\;\;\;\;\;\;\;\;\;
\;\;\;\;\;\;\;\;\;\;\;\;\;\;${\bf 
"Beauty is truth, truth beauty,"-- that is all\\ 
$\;\;\;\;\;\;\;\;\;\;\;\;\;\;\;\;\;\;\;\;\;\;\;\;\;\;\;\;\;\;\;\;\;\;
\;\;\;\;\;\;\;\;\;\;\;\;\;\;\;\;$
Ye know on earth, and all ye need to know.}\\
$\;\;\;\;\;\;\;\;\;\;\;\;\;\;\;\;\;\;\;\;\;\;\;\;\;\;\;\;\;\;
\;\;\;\;\;\;\;\;\;\;\;\;\;\;\;\;\;\;\;\;\;\;\;\;\;\;\;\;\;\;\;\;\;\;
\;\;\;\;\;\;\;\;\;\;\;\;\;\;\;\;\;\;$ 
(John Keats, May 1819)}

\maketitle

\begin{abstract}
\noindent
An updated review \cite{inaugural} of nonextensive statistical 
mechanics and thermodynamics is colloquially presented. Quite 
naturally  the possibility emerges for using the value of $q-1$ 
(entropic nonextensivity) as a simple and efficient manner to 
provide, at least for some classes of systems, some characterization 
of  the degree of what is currently referred to as complexity 
\cite{gellmann} . A few historical digressions are included as well.\\

\end{abstract}


\section{INTRODUCTION}

Would you say that Newtonian mechanics is {\it universal}? We almost 
hear a big and unanimous {\it yes} coming from practically all 
nineteenth-century physicists. And yet, they would all be wrong! 
Indeed, we know today that if the involved masses are very small, say 
that of the electron, Newtonian mechanics has to be replaced by 
quantum mechanics. And if the involved speeds are very high, close to 
that of light in vacuum, special relativity has to be used instead. 
And if the masses are very large, like that of the Sun or of a 
galaxy, general relativity enters into the game. It is our present 
understanding that only in the $(\hbar,\,1/c,\,G) \rightarrow 
(0,\,0,\,0)$ limit, Newtonian mechanics appears to be strictly 
correct. Now, what about Boltzmann-Gibbs statistical mechanics and 
standard thermodynamics? Are they {\it universal}? An attempt to 
answer such question and to clarify its implications constitutes the 
main goal of the present effort. Let us see what Ludwig Boltzmann has 
to tell us about this. In the first page of the second part of his 
{\it Vorlesungen uber Gastheorie}, he qualifies the concept of ideal 
gas by writing: {\it " When the distance at which two gas molecules 
interact with each other noticeably is vanishingly small relative to 
the average distance between a molecule and its nearest neighbor -- 
or, as one can also say, when the space occupied by the molecules (or 
their spheres of action) is negligible compared to the space filled 
by the gas -- ..."}. Were Boltzmann our contemporary, he would have 
perhaps told us that he was addressing systems involving what we 
nowadays call {\it short range} interactions! Several decades ago, 
Laszlo Tisza \cite{tisza}, in his {\it Generalized Thermodynamics}, 
writes {\it ''The situation is different for the additivity 
postulate [...], the validity of which cannot be inferred from 
general 
principles. We have to require that the interaction energy between 
thermodynamic systems be negligible. This assumption is closely 
related 
to the homogeneity postulate [...]. From the molecular point of view, 
additivity and homogeneity can be expected to be reasonable 
approximations for systems containing many particles, provided that 
the 
intramolecular forces have a short range character"}, when referring 
to the usual thermodynamic functions such as internal energy, 
entropy, free energy, and others. Also, Peter T. Landsberg 
\cite{landsbergbook}, in his {\it Statistical Mechanics and 
Thermodynamics}, writes {\it ''The presence of long-range 
forces causes important amendments to thermodynamics, some of which 
are 
not fully investigated as yet."}. More, during a visit to Rio de 
Janeiro some years ago, he told me that {\it such restrictions should 
be in the first pages of {\bf all} books on thermodynamics... but 
they are not!} The title of the 1993 Nature article \cite{cosmology} 
by John Maddox focusing on black holes is {\it ''When entropy does 
not seem extensive"}. What is the viewpoint on such matters of our 
colleagues mathematicians? Distinguished expert in nonlinear 
dynamical systems, Floris Takens writes \cite{takens} {\it ''The 
values of $p_i$ are determined by the following 
dogma: if the energy of the system in the $i^{th}$ state is $E_i$ and 
if the temperature of the system is $T$ then: 
$p_i=\exp\{-E_i/kT\}/Z(T)$, where $Z(T)=\sum_i \exp\{-E_i/kT\}$} 
[...] {\it This choice of $p_i$ is called {\it Gibbs distribution}. 
We shall give no justification for this dogma; even a physicist like 
Ruelle 
disposes of this question as '' deep and incompletely clarified"." } 
It is known that mathematicians sometimes use the word {\it dogma} 
when they do not have the theorem! 
As early as 1964, Michael E. Fisher , and later on with David Ruelle 
as well as with Joel L. Lebowitz, addressed in detail such questions 
\cite{fisherruellelebowitz}. They established, for instance, that a 
$d$-dimensional classical system (say a fluid) including two-body 
interactions that are nonsingular at the origin and decay, at long 
distances, like $r^{-\alpha}$, exhibits standard thermostatistical 
behavior {\it if} $\alpha/d>1$. An interesting question remains open: 
what happens when $\alpha/d \le 1$ ?  This is the case of all 
self-gravitating systems (e.g., galaxies, black holes), a fact which 
explains  well known difficulties in using standard thermodynamics to 
address them. These difficulties are not without relation with
the case of a single hydrogen atom ($d=3$ and $\alpha=1$), for which 
no tractable Boltzmann-Gibbs thermostatistical calculations are 
possible (diverging partition function) unless we confine the 
hydrogen atom in a box \cite{fowler}.  Of course, no such 
difficulties are encountered if we focus on a single harmonic 
oscillator, or on a single spin in an external magnetic field, or on 
a Lennard-Jones fluid ($d=3$ and $\alpha=6$), or even on a neutral 
plasma, where the Coulombian interaction is "dressed" thus becoming a 
thermostatistically innocuous, exponentially decaying interaction. No 
severe anomalies appear in such cases, all relevant sums and 
integrals being finite and perfectly defined. More generally 
speaking, a variety of recent analytical and numerical results (see 
\cite{variety} and references therein) in classical systems (with no 
singularities at the potential origin or elsewhere) suggest that, at 
the appropriately defined ultimate $t \rightarrow \infty$ thermal 
equilibrium, the internal energy {\it per particle} grows, when the 
number of particles $N$ grows, like ${\tilde N} \equiv 
[N^{1-\alpha/d}-\alpha/d] / [1-\alpha/d]$. We easily verify that 
${\tilde N}$ approaches a finite constant if $\alpha/d>1$, diverges 
logarithmically if $\alpha/d=1$, and diverges like $N^{1-\alpha/d}$ 
if $0 \le \alpha/d<1$, thus recovering, for $\alpha=0$, the well 
known Molecular Field Approximation scaling with $N$ (usually 
englobed within the coupling constant whenever dealing with magnetic 
models such as the Ising ones, among many others). We trivially see 
that the standard thermodynamical {\it extensivity} is lost in such 
cases. The physical behavior for the marginal case $\alpha/d=1$ is 
intimately related to the mathematical fact that some relevant sums 
that are {\it absolutely convergent} on the extensive side and {\it 
divergent} on the nonextensive one, become {\it conditionally 
convergent} in that case. This enables us to understand why the 
amount of calories to be provided to a table in order to increase its 
temperature in one degree {\it only} depends on its weight and 
material (iron, wood), whereas the amount of Coulombs we must provide 
to a capacitor to generate a one Volt potential difference at its 
ends {\it also} depends on its {\it shape}! Indeed, the relevant 
interaction in the latter being the permanent dipole-dipole one 
(hence $d=\alpha=3$), the capacity depends on whether the capacitor 
is a slab, round, cylindrical or otherwise. Such behavior 
dramatically contrasts with the extensivity usually focused on in 
textbooks of thermodynamics. 
In fact, it is after all quite natural that self-gravitating systems 
provided the first physical application of the formalism we are 
addressing here. Indeed, physicist Angel Plastino and his son 
astronomer Angel R. Plastino showed in 1993 \cite{plapla} that by 
sufficiently departing from BG thermostatistics, it is possible to 
overcome an old difficulty, namely to have the physically desirable 
feature that total energy, entropy and mass be {\it simultaneously 
finite}.

Let us now try to deepen our quest for understanding the restrictions 
within which Boltzmann-Gibbs statistical mechanics is the appropriate 
formalism for describing a (fantastic!) variety of systems in nature. 
Are long-range interactions the only mechanism which creates 
anomalies? The answer seems to be {\it no}. Indeed, several 
indications exist which suggest that mesoscopic dissipative systems 
(like granular matter) also are thermostatistically anomalous: the 
velocity distribution measured in a variety of computational 
experiments is, as argued by Leo P. Kadanoff et al \cite{kadanoff}, 
Y.-H. Taguchi and H. Takayasu \cite{granular}, Hans J. Herrmann 
\cite{private} and others, far from being Maxwell's Gaussian. On 
general grounds, multifractally structured systems also tend to 
exhibit anomalies. And I believe that it would be surprise for very 
few that the same happened with systems based on strongly 
nonmarkovian microscopic memory (e.g., a memory function decreasing 
towards the past as a slow enough power-law). 

As an attempt to theoretically handle at least some of these 
anomalies, I proposed in 1988 \cite{tsallis} the generalization of 
the Boltzmann-Gibbs formalism by postulating a nonextensive entropy 
$S_q$ (See Sections I and II) which recovers the usual logarithmic 
measure as the $q=1$ particular case. Before going further, let us 
describe $S_q$, its properties and historical context.

\section{\bf THE NONEXTENSIVE ENTROPY $S_q$}

\subsection{Properties}

In the case of a discrete number $W$ of microscopic configurations, 
$S_q$ is given by \cite{tsallis}
\begin{equation}
S_q \equiv k \frac{1-\sum_{i=1}^Wp_i^q}{q-1} \;\;\; 
(\sum_{i=1}^Wp_i=1; \;k>0)
\end{equation}
(similar expressions correspond to the cases where we have a 
continuum of microscopic configurations, or when the system is a 
quantum one).

In the limit $q \to 1$ we have $p_i^{q-1}=e^{(q-1) \ln p_i} \sim 
1+(q-1)\ln p_i$, hence $S_1 = -k\sum_{i=1}^W p_i\;\ln p_i$, which is 
the usual Boltzmann-Gibbs-Shannon expression. Also, it can be shown 
that 

(i) $S_q \ge 0$, and, for $q>0$, equals zero in the case of certainty 
(i.e., when all probabilities vanish but one which equals unity);

(ii) $S_q$ attains its extremum (maximum for $q>0$ and minimum for 
$q<0$) for equiprobability (i.e., $p_i=1/W$), thus becoming $S_q=k 
[W^{1-q}-1]/[1-q]$. This expression becomes, for $q=1$, $S_1=k \ln 
W$, which is the formula graved on Boltzmann's tombstone in the 
Central Cemetery in Vienna;

(iii) $S_q$ is concave (convex) for $q>0$ ($q<0$), which is the basis 
for thermodynamic stability;

(iv) If two systems $A$ and $B$ are independent in the sense of the 
theory of probabilities (i.e., $p_{ij}^{A+B} = p_i^A \times p_j^B$), 
then
\begin{equation}
\frac{S_q(A+B)}{k}= 
\frac{S_q(A)}{k}+\frac{S_q(B)}{k}+(1-q)\frac{S_q(A)}{k}\frac{S_q(B)}{k}\;,
\end{equation}
hence, superextensivity, extensivity or subextensivity occurs when 
$q<1$, $q=1$ or $q>1$ respectively;

(v) If the $W$ states of a system are divided into $W_L$ and $W_M$ 
states ($W_L +W_M=W$), and we define $p_L= \sum_{W_L states}p_l$ and 
$p_M=\sum_{W_M states}p_m$ ($p_L+p_M=1$), then
\begin{equation}
S_q(\{p_i\})=S_q(p_L,p_M)     + p_L^q 
S_q(\{p_l/p_L\})+p_M^qS_q(\{p_m/p_M\})
\end{equation}
which generalizes Shannon's celebrated formula;

(vi) It can be easily checked that $-d (\sum_i 
p_i^x)/dx|_{x=1}=-\sum_ip_i \ln p_i$. Sumiyoshi Abe, in Funabashi, 
made in 1997 \cite{sumiyoshi} a simple but very deep remark. He 
considered the so-called {\it Jackson derivative}. The mathematician 
F. Jackson generalized, one century ago, the usual differential 
operator. He defined $D_qf(x) \equiv [f(qx)-f(x)]/[qx-q]$, which 
reproduces $d/dx$ in the limit $q \rightarrow 1$. Abe' s observation 
consists in the following easily verifiable property, namely 
$-D_q(\sum_i p_i^x)/dx|_{x=1}=S_q$ ! Since it is clear that Jackson 
derivative "tests" how the function $f(x)$ "reacts" under dilatation 
of $x$ in very much the same way the standard derivative "tests" it 
under translation of $x$, Abe's remark opens a wide door to physical 
insight onto thermodynamics. It was so perceived, I believe, by 
Murray Gell-Mann, of the Santa Fe Institute. Indeed, one year ago, 
during a conference in Italy, he asked me whether I would go to 
conjecture that systems with relevant symmetries {\it other} than say 
translation or dilatation invariances would need entropies {\it 
other} than $S_1$ or $S_q$. My answer was {\it probably yes}. Indeed, 
there are symmetries that appear in general relativity or in string 
theory that are characterized by a sensible amount of parameters. The 
concept of thermodynamic information upon such systems, i.e., the 
appropriate entropy could well be a form involving essentially a 
similar number of parameters. {\it If} so, then symmetry would 
control entropy, and $S_q$ would only be the beginning of a 
presumably long and fascinating story! Since we have learnt from 
Gell-Mann and others, how deeply symmetry controls energy, we could 
then say that symmetry controls thermodynamics, that science which 
nothing is but a delicate balance between energy end entropy. More 
precisely, symmetry would then determine (see also \cite{reichl}) the 
{\it specific microscopic form} that Clausius thermodynamic entropy 
would take for {\it specific} classes of systems. Symmetry 
controlling thermodynamics! If such a conjectural scenario turned out 
to be correct, this would not be to displease Plato with his unified 
view of truth and beauty!

(vii) Along the lines of Claude E. Shannon's theorem, it has been 
shown by Roberto J. V. dos Santos, in Maceio, Brazil \cite{santos}, 
that a set of conditions generalizing (through properties (iv) and 
(v)) that imposed by Shannon uniquely determines $S_q$. Analogously, 
Abe recently showed \cite{abeuniqueness} that consistently 
generalizing Khinchin's set of conditions, once again a unique 
solution emerges, namely $S_q$.

\subsection{\bf Brief review of the labyrinthic history of the 
entropies}
 
The entropy and the entropic forms have haunted physicists, chemists, 
mathematicians, engineers and others since long! At least since 
Clausius coined the word and wrote {\it ''The energy of the universe 
is constant, its entropy tends to increase"} (See Appendix). The 
whole story is a labyrinth plenty of rediscoveries. After Ludwig 
Boltzmann and Josiah Willard Gibbs introduced and first analyzed, 
more than one century ago, their respective expressions 
\cite{lebowitz} of the entropy in terms of microscopic quantities, 
many generalizations and reformulations have been proposed and used. 
Let us mention some of the most relevant ones. John von Neumann 
introduced a quantum expression for the entropy in terms of the 
density operator; when the operator is diagonalized, the traditional 
expression (herein referred, because of the functional form, to as 
the Boltzmann-Gibbs entropy, although these two scientists used to 
work with different, though consistent, expressions \cite{lebowitz}) 
is recovered. Shannon reinvented, more than half a century ago, a 
binary version of the B-G functional form and interpreted it in terms 
of information and communication theory. It is no doubt very  
interesting to notice that in {\it The mathematical theory of 
communication} he wrote, in reference to the specific form he was 
using, that {\it "This theorem, and the assumptions required for its 
proof, are in no way necessary for the present theory. It is given 
chiefly to lend a certain plausibility to some of our later 
definitions. The real justification of these definitions, however, 
will reside in their implications."}. In 1957, Edwin Jaynes 
attributed to the whole formalism a very generic, and still 
controversial in spite of an enormous amount of practical 
applications, anthropomorphic shape. Since those years, people 
working in cybernetics, information theory, complexity and nonlinear 
dynamical systems, among others, have introduced close to 20 
different entropic forms! This number has almost doubled, this time 
mainly because of physicists, since the 1988 paper \cite{tsallis} 
appeared. 

In 1960, A. Renyi proposed \cite{renyi} a form which recovers that of 
Shannon as a special case. His form is {\it always extensive}, but 
{\it not always concave (or convex)} with regard to the probability 
distributions. It seems that it was a rediscovery; indeed, according 
to  I. Csiszar's 1974 critical survey \cite{beforerenyi}, that form 
had essentially already been introduced by Paul-Marcel Schutzenberger 
in 1954 \cite{schutzenberger}. In 1967, presumably for cybernetic 
goals, J. Harvda and F. Charvat introduced \cite{harvda}, although 
with a different prefactor (fitted for binary variables), the 
entropic form herein noted $S_q$, and studied some of its 
mathematical properties. One year later, I. Vajda further studied 
\cite{vajda}this form, quoting Harvda and Charvat. In 1970, in the 
context of information and control, Z. Daroczy rediscovered 
\cite{daroczy} this form (he quotes neither Harvda and Charvat, nor 
Vajda). Later on, in 1975, B.D. Sharma and D.P. Mittal  introduced 
\cite{sharma} and studied some mathematical properties of a 
two-parameter form which recovers {\it both} Renyi's and $S_q$ as 
particular instances! In his 1978 paper in Reviews of Modern Physics, 
A. Wehrl mentions \cite{generalentropies} the form of $S_q$, quoting 
Daroczy, but not Harvda, Charvat, Vajda, Sharma and Mittal. In 1988, 
in the Physics community, a new rediscovery, by myself this time. 
Indeed, inspired by multifractals, I postulated \cite{tsallis} the 
form $S_q$ as a possible new path for generalizing Boltzmann-Gibbs 
statistical mechanics; through optimization of $S_q$ I obtained an 
equilibrium-like distribution which generically is a {\it power-law}, 
and reproduces Boltzmann's celebrated factor as the particular limit 
$q \rightarrow 1$. In my 1988 paper, I quoted only Renyi's entropy, 
having never taken notice of all these generalized forms, whose 
existence (and that of many others!) I have been discovering along 
the years. In fact, Renyi's entropy can be shown \cite{tsallis} to 
be  a monotonically increasing function of $S_q$. However, concavity 
is not preserved through monotonicity. So what?, since the optimizing 
distribution 
will be exactly the same. Well, it happens that statistical mechanics 
is much more than an equilibrium-like distribution optimizing an 
entropy 
under some generic constraints! It includes a variety of other 
relevant points, such as the role played by fluctuations, linear 
responses, thermodynamical stability, to name but a few. Nobody 
would, I believe, try to reformulate BG statistical mechanics using, 
instead of the BG entropy, say its {\it cube}, just because it is a 
monotonically increasing function of the BG one! Let us also mention 
that, within various scientific and technological communities, close 
to 30 (!) different entropic forms have been considered along the 
last half a century. 

Let us summarize some of what we have said. The BG entropy, $S_1$, is 
both {\it concave} and {\it extensive}, the Renyi entropy, $S_q^R$, 
is {\it extensive} but {\it nonconcave} (for all $q>0$; concave only 
for $0<q \le 1$), the one focused on herein, $S_q$, is {\it 
nonextensive} but {\it concave} (for all $q>0$), and the 
Sharma-Mittal entropy is {\it nonextensive} and {\it nonconcave}. The 
latter is determined by {\it two} parameters and contains the other 
three as particular cases. The Renyi entropy and $S_q$ are determined 
by {\it one} parameter ($q$) and share a common case, namely the BG 
entropy, which is nonparametric. Since Renyi's, Sharma-Mittal's and 
the present entropy depend from the probability set through one and 
the same expression, namely $\sum_i p_i^q$, they are all three 
related through simple functions. In particular,
\begin{equation}
S_q^R \equiv k\frac{\ln \sum_i p_i^q}{1-q} = 
k\frac{\ln[1+(1-q)S_q/k]}{1-q}\;\;\;(\forall q),
\end{equation}
with $S_1^R=S_1=-k\sum_i p_i \ln p_i$.

For completeness, let us close this section by mentioning that 
expressions of the type $\sum_i p_i^q$ have also been, long ago, 
discussed on mathematical grounds by Hardy, Littlewood and Polya 
\cite{hardyetal} (and, for $q=2$, by the Pythagoreans!).

\section{Equilibrium-like statistics}

The next step is to follow along Gibbs path. For instance, to 
formally obtain the "equilibrium" distribution associated with the 
canonical ensemble (i.e., a conservative system at "equilibrium" with 
a thermostat, where the physical meaning of this state will turn out 
to be, as we shall see later on, close to a stationary one), we need 
to impose a constraint on total energy. We shall adopt 
\cite{curadotsamepla} that $\langle {\cal H} \rangle_q \equiv 
\sum_iP_i \epsilon_i = U_q$, where $U_q$ is fixed and finite, 
$\{\epsilon_i\}$ is the set of energy levels of the system 
Hamiltonian ${\cal H}$, and the {\it escort} distribution $\{P_i\}$ 
is defined as $P_i \equiv p_i^q/[\sum_j p_j^q]$; from now on we refer 
to $\langle ...\rangle_q$ as the {\it normalized $q$-expectation 
value}. At the present stage, a plethora of mathematical reasons 
exist for adopting this particular generalization of the familiar 
concept of {\it internal energy}. Not the least of these reasons is 
the fact that such definition enables relevant sums and integrals to 
be {\it finite}, which would otherwise diverge (e.g., the second 
moment of L\'evy distributions diverges, whereas the second 
$q$-moment with $q$ appropriately chosen is {\it finite}). We can say 
that in this sense the theory becomes {\it regularized}. The 
optimization of $S_q$ with this constraint (to which we associate a 
Lagrange parameter $\beta$) leads to 
\begin{equation}
p_i=[1-(1-q) \beta_q (\epsilon_i-U_q)]^{1/(1-q)}/Z_q\;\;(\beta_q 
\equiv \beta / Z_q^{1-q})
\end{equation}
with $Z_q \equiv \sum_j [1-(1-q) \beta_q (\epsilon_j-U_q)]^{1/(1-q)}$ 
(See Fig. 1). We easily verify that $q \rightarrow 1$ recovers the 
celebrated, {\it exponential} Boltzmann factor
\begin{equation}
p_i=e^{-\beta \epsilon_i}/Z_1 \;\;\;(Z_1 \equiv \sum_je^{-\beta 
\epsilon_j}).
\end{equation}
 For $q>1$ a {\it power-law} tail emerges; for $q<1$ the formalism 
imposes a high-energy {\it cutoff}, i.e., $p_i=0$ whenever the 
argument of the power function becomes negative. One comment is 
worthy: this distribution is generically {\it not} an exponential 
law, i.e, it is generically {\it not} factorizable (under sum in the 
argument), and nevertheless is {\it invariant} under the choice of 
zero energy for the energy spectrum! (this is in fact one of the 
aside benefits of defining the constraints in terms of {\it 
normalized} distributions like the escort ones). 

Eq. (5) can be rewritten in the following convenient form:
\begin{equation}
p_i=[1-(1-q) \beta^{\prime} \epsilon_i]^{1/(1-q)}/Z_q^{\prime}
\end{equation}
with $Z_q^{\prime} \equiv \sum_j [1-(1-q) \beta^{\prime} 
\epsilon_j]^{1/(1-q)}$, where $\beta^{\prime}$ is a simple function 
of $\beta_q$ and $U_q$.

It can be shown \cite{curadotsamepla}, along the lines of the 1991 
paper with Evaldo M.F. Curado \cite{curadotsamepla}, that this 
formalism satisfies, for {\it arbitrary} $q$, $1/T=\partial S_q / 
\partial U_q$ ($T \equiv 1/(k \beta)$), and the entire Legendre 
transform structure of standard thermodynamics. Moreover, the 
relations which provide $U_q$ and $F_q=U_q-TS_q$ in terms of $Z_q$ 
(i.e., the connection between the microscopic and macroscopic 
descriptions of the system), remain basically the {\it same} as the 
usual ones, the role played by the logarithmic function $\ln x$ being 
now played by $(x^{1-q}-1)/(1-q)$. Finally, various important 
thermostatistical theorems can be shown to be $q$-invariant. Among 
them, we must mention the Boltzmann H-theorem (macroscopic time 
irreversibility, i.e., essentially the second principle of 
thermodynamics, first tackled within this context by Ananias M. Mariz 
in Natal, Brazil \cite{mariz}), Ehrenfest theorem (correspondence 
principle between classical and quantum mechanics), Einstein 1910 
factorizability of the likelihood function in terms of the entropy 
(this $q$-generalization was first realized in 1993 during a private 
discussion with Manuel O. Caceres, from Bariloche, Argentina), 
Onsager reciprocity theorem (microscopic time reversibility), Kramers 
and Wannier relations (causality), and, apparently, even Pesin 
relation for nonlinear dynamical systems, although the last one 
remains at the level of plausibility (no proof available). The 
$q$-generalization of the fluctuation-dissipation theorem was first 
addressed by Anna Chame and Evandro V.L. de Mello, in Niteroi, Brazil 
\cite{chame}, and then, along with the generalization of Kubo's 
linear response theory, by A.K. Rajagopal, in the Naval Research 
Laboratory in Washington \cite{rajagopal}. Since then, a variety of 
important statistical mechanical approaches such as the Bogolyubov 
inequality, the many-body Green function and path integral formalisms 
have been generalized by Renio S. Mendes in Maringa, Brazil, Ervin K. 
Lenzi in Rio de Janeiro, and collaborators \cite{rajagopal}. 
Nonextensive versions of Fermi-Dirac and Bose-Einstein statistics 
have been discussed \cite{buyukkilic} by Fevzi Buyukkilic, Dogan 
Demirhan and Ugur Tirnakli in Izmir, Turkey, and their collaborators. 
Finally, it is certainly important to mention that, very recently, 
Rajagopal and Abe have done an exhaustive review \cite{rajabebalian} 
of apparently {\it all} the traditional methods for obtaining the BG 
equilibrium distribution, namely the Balian-Balazs counting method 
within a microcanonical basis, the Darwin-Fowler steepest descent 
method, and the Khinchin's large-numbers method. They $q$-generalized 
all three, and systematically obtained the same power-law 
distribution obtained \cite{tsallis,curadotsamepla} above through the 
Gibbs' method, i.e.,  the optimization of an appropriately 
constrained entropy.

In order to see the nonextensive formalism at work, let us address 
\cite{levy} an important question which long remained without 
satisfactory solution, namely, why L\'evy distributions are 
ubiquitous in nature, in a manner which by all means is similar to 
the ubiquity of Gaussians?  To introduce the point, let us first 
address the Gaussian case. If we  optimize the BG entropy $S_1$ by 
imposing a fixed and {\it finite} average $\langle x^2\rangle$, we 
obtain a Gaussian one-jump distribution. By convoluting $N$ such 
distributions we obtain a $N$-jump distribution which also is a 
Gaussian. From this immediately follows that $\langle x^2\rangle(N) 
\propto N$. If we consider that $N$ is proportional to time, we have 
the central result of Einstein's 1905 celebrated paper on Brownian 
motion, which provided at the time strong evidence in favor of 
Boltzmann's ideas. We may reword this result by saying that the 
foundation of Gaussians in nature lies upon two pillars, namely the 
BG entropy and the central limit theorem. What about L\'evy 
distributions? The difficulty comes from the fact that all L\'evy 
distributions have {\it infinite} second moment, due to their long 
tail ($\propto 1/|x|^{1+\gamma}$ with $0<\gamma<2$).  Consequently, 
what simple auxiliary constraint to use along with the optimization 
of the entropy? In 1994, Pablo Alemany and Damian H. Zanette in 
Bariloche, Argentina, settled \cite{alemany} the basis for the 
solution, namely optimization of $S_q$ while imposing a {\it finite} 
$q$-expectation value of $x^2$. The one-jump distribution thus 
obtained is proportional to $[1-(1-q)\beta x^2]^{1/(1-q)}$, which 
respectively reproduces the Gaussian, Lorentzian ({\it Cauchy} for 
the mathematicians) and completely flat distributions for $q=1,\;2$ 
and $3$. For $q<1$ the support is compact, and for $q>1$ a  power-law 
tail is obtained. The proof was completed one year later \cite{levy}, 
when it was argued that the L\'evy-Gnedenko central limit theorem was 
applicable for $q>5/3$, thus approaching, for $N>>1$ precisely L\'evy 
distributions for the $N$-jump distribution. We may then summarize 
this result by saying that the foundation of L\'evy distributions in 
nature also has two pillars, namely $S_q$ and the L\'evy-Gnedenko 
theorem. It follows that $\gamma=2$ for $q \le 5/3$ and $\gamma = 
(3-q)/(q-1)$ for $5/3 <q<3$. Since $\gamma$ can also be interpreted 
as the fractal dimension to be associated with L\'evy walks, this 
turned out to be the first {\it exact} connection between 
nonextensive statistical mechanics and scaling. Similar ideas have 
been successfully put forward by Plastino and Plastino \cite{bukman} 
and by Lisa Borland \cite{borland} and others, concerning nonlinear 
as well as fractional-derivative Fokker-Planck-like equations. In all 
these various types of anomalous diffusion, the Gaussian solution 
associated with Jean Baptiste Joseph Fourier's bicentennial heat 
equation is recovered as the $q=1$ limit. Let me also mention that 
Hermann Haken, in Stuttgart, and collaborators have recently applied 
related ideas within a theoretical model of the human brain 
\cite{haken}. 

In the same vein which led to the above discussion of L\'evy 
distributions, if we optimize $S_q$ imposing, besides normalization 
of $p(x)$, finite values for the normalized $q$-expectation values of 
{\it both} $x$ and $x^2$, we obtain
\begin{equation}
p_q(x) \propto [1-(1-q)(\beta^{(1)}x + \beta^{(2)}x^2)]^{1/(1-q)}\;,
\end{equation}
where $\beta^{(1)}$ and $\beta^{(2)}$ are determined by the 
constraints. This distribution contains, as particular cases, a 
considerable amount of well (and not so well) known distributions, 
such as the exponential, the Gaussian, the Cauchy-Lorentz, the 
Edgeworth \cite{edgeworth}(for $\beta^{(1)}=1$, $\beta^{(2)}=0$, and 
$x \ge 0$), the $r$- and the Student's $s$- ones \cite{andre}, among 
others. Let us end with a short and amusing historical remark 
concerning that which can be considered as the most famous of all 
nontrivial distributions, namely the normal one. In contrast with 
what almost everybody would naively think, it was \cite{stigler} 
first introduced by Abraham De Moivre in 1733, then by Pierre Simon 
de Laplace in 1774, then by Robert Adrain in 1808, and finally by 
Carl Friedrich Gauss in 1809, nothing less than 76 years after its 
first publication!

\section{Microscopic dynamical foundation}

To have a deeper understanding of what this generalized formalism 
means, it is highly convenient to focus on the mixing properties.

\subsection{\bf Mixing in one-dimensional dissipative maps}

Let us now address a very simple type of system, which will 
nevertheless provide important hints about the physical and 
mathematical significance of the present formalism. Let us consider 
the well known logistic map, namely
\begin{equation}
x_{t+1} = 1 -a x_t^2\;\;\;(t=0,\;1,\;2,\;...;\; 0 \le a \le 2)
\end{equation}
For $a<a_c = 1.40115519...$, the attractors are {\it finite} cycles 
(fixed points, cycle-2, cycle-4, etc) and the Lyapunov exponent 
(hereafter noted $\lambda_1$) is almost everywhere {\it negative} 
(i.e., {\it strongly insensitive} to the initial conditions and any 
intermediate rounding).  For $a_c \le a \le 2$, the attractor is 
chaotic for most values (i.e., it has an {\it infinite} number of 
elements, analogously to an irrational number) and the $\lambda_1$ 
exponent is consistently {\it positive} (i.e., {\it strongly 
sensitive} to the initial conditions and any intermediate rounding). 
There is however, in the interval $[0,\;2]$, an infinite number of 
values of $a$ for which $\lambda_1=0$. What happens with the 
sensitivity to the initial conditions in those marginal cases? Let us 
focus on this problem by first recalling the definition of 
$\lambda_1$. If two initial conditions $x_0$ and $x_0^{\prime}$ 
differ by the small amount $\Delta x_0$, we can follow the time 
evolution of $\Delta x_t$ through the quantity $\xi \equiv 
\lim_{\Delta x_0 \rightarrow 0} (\Delta x_t/\Delta x_0)$. The most 
frequent situation is that $\xi = e^{\lambda_1\;t}$, which defines 
the Lyapunov exponent $\lambda_1$. We can trivially check that 
$\xi(t)$ is the solution of $d\xi / dt=\lambda_1 \;\xi$. We have 
conjectured \cite{logistic} that, whenever $\lambda_1$ {\it vanishes} 
for this map (and similar ones), the controlling equation becomes 
$d\xi / dt=\lambda_q \;\xi^q$, hence $\xi = 
[1+(1-q)\lambda_q\;t]^{1/(1-q)}$. This result recovers the usual one 
at the $q=1$ limit; also, it defines a {\it  generalized} Lyapunov 
coefficient $\lambda_q$, which, like $\lambda_1$, inversely scales 
with time, but now within a law which generically is a {\it power}, 
instead of the standard {\it exponential}. The two paradigmatic cases 
are $q<1$ (with $\lambda_q>0$), hereafter referred to as {\it weakly 
sensitive} to the initial conditions, and $q>1$ (with $\lambda_q<0$), 
hereafter referred to as {\it weakly insensitive} to the initial 
conditions. Within this unified scenario, we would have {\it strong 
chaos} for $q=1$  and $\lambda_1>0$ (i.e., {\it exponential mixing}) 
and {\it weak chaos} for $q<1$, $\lambda_1=0$ and $\lambda_q>0$ 
(i.e., {\it power-law mixing}). All these various possibilities have 
indeed been observed for the logistic map. If we start with $x_0=0$, 
we have (i) $\lambda_1<0$ for $a=a_c-10^{-3}$, (ii) $\lambda_1>0$  
for $a=a_c+10^{-3}$, (iii) $\lambda_1=0$, $q>1$ and $\lambda_q<0$ for 
a doubling-period bifurcation (e.g., $a=3/4$) as well as for a 
tangent bifurcation (e.g., $a=7/4$), and finally (iv) $\lambda_1=0$, 
$q<1$ and $\lambda_q>0$ at the edge of chaos (i.e., $a=a_c$). The 
last case is, for the present purposes, the most interesting one. 
This power-law mixing was first observed and analyzed by P. 
Grassberger, A. Politi, H. Mori and collaborators \cite{grassberger}. 
Its importance seems to come mainly from the fact that, in the view 
of many authors, "life appeared at the edge of chaos"! Moreover, this 
type of situation seems to be a paradigm for vast classes of the 
so-called {\it complex systems}, much in vogue nowadays. Let us say 
more about this interesting point. The precise power-law function 
$\xi$ indicated above appears to be the {\it upper bound} of a 
complex time-dependence of the sensitivity to the initial conditions, 
which includes considerable and ever lasting fluctuations. From a 
log-log representation we can numerically get the slope and, since 
this is to be identified with $1/(1-q)$, we obtain the corresponding 
value $q^*$. For the logistic map we have obtained \cite{logistic} 
$q^*=0.24...$. If we apply these concepts to a more general map, like 
$x_{t+1} = 1 -a |x_t|^z\;\;\;(0 \le a \le 2; \; z>1)$, we can check, 
and this is long known, that the edge of chaos depends on $z$. More 
precisely, when $z$ increases from 1 to infinity, $a_c$ increases 
from 1 to 2. If for $a_c$ we check the sensitivity to the initial 
conditions, we obtain that $q^*$ increases from $-\infty$ to close to 
1 (though smaller than 1), when $z$ increases from 1 to infinity. 
Simultaneously, the fractal dimension $d_f$ of the chaotic attractor 
increases from 0 to close to 1(though smaller than 1). If instead of 
the logistic family of maps, we use some specific family of circle 
maps also characterized by an inflexion exponent $z$, it can be shown 
that, once again, $q^*$ increases with $z$ for all studied values of 
$z$ ($z \ge 3$). The interesting feature in this case is that $d_f=1$ 
for {\it all} these values of $z$. What we learn from this is that 
the index $q$ characterizes essentially the {\it speed} of the 
mixing, and only indirectly how "filled" is the phase space within 
which this mixing is occuring.

Let us now address, for maps like the logistic one, a {\it second} 
manner of obtaining $q^*$ , this time based on the multifractal 
geometry of the chaotic attractor. This geometry can be, and usually 
is, characterized by the so called $f(\alpha)$ function, a down-ward 
parabola-like curve which generically is below (and tangential to) 
the bissector in the $(\alpha,f)$ space, is concave, its maximal 
value is $d_f$, and, in most of the cases, vanishes at two points, 
namely $\alpha_{min}$ and $\alpha_{max}$ (typically, $0 < 
\alpha_{min} <d_f<\alpha_{max}$). For a fractal, one expects, 
$\alpha_{min} 
=d_f=\alpha_{max}=f(\alpha_{min})=f(d_f)=f(\alpha_{max})$. For a 
so-called {\it fat} (multi)fractal, $d_f$ attains the value of the 
Euclidean space within which the (multi)fractal is embedded (this is 
the case of the family of circle maps mentioned above). Marcelo L. 
Lyra, in Maceio, and myself used (imitating the successful Ben Widom 
style!) in 1998 \cite{logistic} some simple scaling arguments  and 
obtained the following relation:
\begin{equation}
\frac{1}{1-q^*}=\frac{1}{\alpha_{min}}- 
\frac{1}{\alpha_{max}}\;\;\;(q^*<1)
\end{equation}
This is a kind of fascinating relation since it connects  the 
power-law sensitivity of nonlinear dynamical systems with purely 
geometrical quantities. It has been numerically checked for at least 
three different one-dimensional dissipative maps. Its analysis for 
two- or more- dimensional systems, as well as for conservative ones, 
would of course be very  enlightening and remains to be done (as do 
quantum analogs of this problem). Furthermore, it does not contain in 
a generic way the $q=1$ limit. Indeed, it seems that, for its 
validity, it is implicitly assumed that 
$f(\alpha_{min})=f(\alpha_{max})=0$. A more general expression could  
be something like
$1/(1-q^*)= 1/[\alpha_{min} - f(\alpha_{min})] - 1/[\alpha_{max} - 
f(\alpha_{max})]$; the generic $q=1$ limit could then correspond to 
$\alpha_{min}=f(\alpha_{min})$. The second manner we referred above 
goes as follows. We can numerically construct the $f(\alpha)$ 
function associated with the chaotic attractor, measure 
$\alpha_{min}$ and $\alpha_{max}$, and, through the above connection, 
calculate $q^*$. This procedure yields, for the standard logistic 
map, $q^*=0.2445...$, thus reproducing the number we previously 
obtained from the sensitivity to the initial conditions. The higher 
precision we indicate here comes from the fact that, for the 
logistic-like family  maps, it is long known that 
$\alpha_{max}=z\;\alpha_{min}= \frac{\ln 2}{\ln  \alpha_F}$, where 
$\alpha_F$ denotes the corresponding Feigenbaum constant. This 
constant being known with many digits for $z=2$, we can infer $q^*$ 
with quite high precision. 

Let us finally address now a {\it third} manner of obtaining $q^*$, 
this time directly from the nonextensive entropy $S_q$. It is known 
that, for large classes of nonlinear dynamical systems, the rate of 
loss of information as time increases can be characterized
by the so called Kolmogorov-Sinai entropy (hereafter noted $K_1$). 
This quantity is defined, strictly speaking, through a {\it 
single-trajectory} procedure, based on the frequencies of appearance, 
in increasingly long strips, of symbolic sequences associated with 
partitions of the phase space. However, in apparently very large (if 
not all) classes of problems, this single-trajectory procedure can be 
replaced by an {\it ensemble-based} one, whose numerical 
implementation is by far simpler computationally than the 
implementation of algorithms based on sequences. We shall restrict 
ourselves to this ensemble-based procedure. Let us illustrate, on the 
logistic map, the steps to be followed in order to calculate $K_1$. 
We choose a value of $a$ for which $\lambda_1$ is positive, say $a=2$ 
(whose corresponding Lyapunov exponent is $\lambda_1= \ln2$). We then 
partition the accessible phase space ($-1 \le x \le 1$, in our 
example) in $W>>1$ little windows, and choose (in any way), inside 
one of those windows, $N$ initial values for $x_0$. We then apply the 
map onto each of those initial conditions, which, as time evolves, 
will start spreading within the accessible phase space. At time $t$, 
we will observe a set $\{N_i(t)\}$ of points inside the $W$ windows 
($\sum_{i=1}^W N_i(t) = N,\;\forall t$). We next define a set of 
probabilities $p_i \equiv N_i/N\;(i=1,\;2,\;...,\;W)$, which enable 
in turn the calculation of $S_1(t) =-\sum_{i=1}^Wp_i\; \ln p_i$. This 
entropy starts, by construction, from zero at $t=0$, and tends to 
linearly increase as time goes on, finally saturating at a value 
which, of course, cannot exceed $\ln W$. The Kolmogorov-Sinai-like 
entropy rate is then defined through $K_1 \equiv \lim_{t \rightarrow 
\infty} \lim_{W \rightarrow \infty} \lim_{N \rightarrow \infty} 
S_1(t)/t$. We expect, according to the Pesin equality, 
$K_1=\lambda_1$. Indeed, we numerically obtain $K_1 = \ln2$ (see Fig. 
2(a)). We can analogously define a generalized quantity, namely 
$K_q\equiv \lim_{t \rightarrow \infty} \lim_{W \rightarrow \infty} 
\lim_{\rightarrow \infty} S_q(t)/t$. And we can numerically verify 
that, for say $a=2$, $K_q$ vanishes (diverges) for any value of $q>1$ 
($q<1$), {\it being finite only for $q=1$}. Let us now turn onto the 
edge of chaos, i.e., $a=a_c$. Vito Latora and Andrea Rapisarda, in 
Catania, Italy, Michel Baranger, at MIT, and myself \cite{logistic}, 
have numerically verified the same structure, excepting for the fact 
that now $K_q$ vanishes (diverges) for any value $q>q^*$ ($q<q^*$), 
{\it being finite only for $q=q^*=0.24...$} ! (see Fig. 2(b)). 
Tirnakli et al \cite{garin} have also verified the validity of this 
structure for various typical values of $z$ within the logistic-like 
family of maps. 

Let us summarize this section by emphasizing that we have seen that 
it is possible to arrive to one and the same nontrivial value of 
$q^*$ through {\it three} different and suggestive roads, namely 
sensitivity to the initial conditions, multifractal structure, and 
rate of loss of information. By so doing, we obtain insight onto the 
mathematical and physical meaning of $q$ and its associated 
nonextensive formalism. Moreover, a pleasant unification is obtained 
with the concepts already known for standard, exponential mixing. A 
very interesting question remains open though, namely the 
generalization of Pesin equality. It might well be  that, as 
conjectured in \cite{logistic}, $K_q$ equals $\lambda_q$ (or, more 
generically, an appropriate average of $\lambda_q(x)$, whenever 
nonuniformity is present). However, a mathematical basis for 
properties such as this one is heavily needed. Some recent numerical 
results along these lines by Paolo Grigolini, in Denton, Texas, and 
collaborators \cite{grigolini} seem promising. In the same vein, it 
is known that whenever {\it escape} exists, as time evolves, out of a 
specific region of the phase space of a chaotic system, $K_1$ becomes 
larger than $\lambda_1$, the difference being precisely the {\it 
escape rate} \cite{dorfman}. All these quantities are defined through 
{\it exponential} time-dependences. What happens if all three are 
{\it power} laws instead? A natural speculation emerges, namely that 
perhaps a similar relation exists for $q \ne 1$.

\subsection{\bf Mixing in many-body  dissipative maps}

In the previous section we described a variety of interesting 
nonextensive connections occurring in one-dimensional dissipative 
systems. A natural subsequent question then is what happens when the 
dissipative system has many, say $N>>1$, degrees of freedom? Such is 
the case of essentially all the models exhibiting Per Bak's 
self-organized criticality (SOC). Francisco Tamarit and Sergio 
Cannas, in Cordoba, Argentina, and myself have discussed 
\cite{evolute}this issue for the Bak-Sneppen model \cite{baksneppen} 
for biological evolution. By numerically studying the time evolution 
of the Hamming distance for increasingly large systems, a power-law 
was exhibited. By identifying the exponent with $1/(1-q^*)$, the 
value $q^* \simeq -2.1$ was obtained. Totally analogous results are 
obtained for the Suzuki-Kaneko model for imitation games (e.g., bird 
singings) \cite{pajaros}. It would be of appreciable importance if 
the same numbers were achieved through the other two procedures, 
namely the determination of the $f(\alpha)$ function and of the loss 
of information. Such consistency would be of great help for further 
understanding. These tasks remain, however, to be done. 

\subsection{\bf Mixing in many-body Hamiltonian systems}

Let us know address what might be considered as the {\it Sancta 
Sanctorum} of statistical mechanics, namely the systems that 
Boltzmann himself was mainly focusing on, i.e., classical many-body 
Hamiltonian systems. We know that no severe thermostatistical 
anomalies arrive for $N$-body $d$-dimensional systems whose (say 
two-body ) interactions are neither singular at the origin nor 
long-ranged. A most important example which violates {\it both} 
restrictions is of course Newtonian gravitation. We shall here 
restrict to a simpler case, namely interactions that are well behaved 
at the origin, but which can be long-ranged. For specificity, let us 
assume a two-body potential which (attractively) decays at long 
distances like $1/r^{\alpha}$ with $0 \le \alpha$ (for historical 
reasons we use $\alpha$ for this exponent; however, it has clearly 
nothing to do with the $\alpha$ appearing in the characterization of 
multifractals previously discussed). The limit $\alpha \rightarrow 
\infty$ corresponds to very short interactions; the other extreme 
($\alpha=0$) corresponds to a typical Mean Field Approximation. No 
thermostatistical basic difficulties are expected for 
$\alpha>\alpha_c$ but the situation is quite different for $0 \le 
\alpha \le \alpha_c$. As already mentioned, Fisher and collaborators 
\cite{fisherruellelebowitz} established that $\alpha_c=d$ for 
classical systems ($\alpha_c \ge d$ is expected for quantum systems). 
Therefore, the nontrivial case is $0 \le \alpha/d \le 1$. Two 
physically different situations are to be understood, namely what 
happens at the $\lim_{N \rightarrow \infty} \lim_{t \rightarrow 
\infty}$ situation, and what happens at the  $\lim_{t \rightarrow 
\infty} \lim_{N \rightarrow \infty} $ one. The two orderings are 
expected to be essentially equivalent for $\alpha/d >1$, and 
dramatically different otherwise. Let us first focus on the $\lim_{N 
\rightarrow \infty} \lim_{t \rightarrow \infty}$ ordering. This 
should correspond to the traditional, BG-like, concept of thermal 
equilibrium, though with anomalous scalings. Let us be more specific 
and consider a fluid thermodynamically described by
\begin{equation}
G(T,p,N)=U(T,p,N) - T S(T,p,N) + p V(T,p,N),
\end{equation}
hence, using the quantity ${\tilde N}$ introduced in Section I,
\begin{equation}
\frac{G(T,p,N)}{N{\tilde N}}=\frac{U(T,p,N)}{N{\tilde N}} - 
\frac{T}{{\tilde N}} \;\frac{S(T,p,N)}{N} + \frac{p}{{\tilde N}}\; 
\frac{V(T,p,N)}{N}\;.
\end{equation}
In the thermodynamic limit $N \rightarrow \infty$ we expect to obtain 
{\it finite} quantities as follows:
\begin{equation}
g({\tilde T},{\tilde p})=u({\tilde T},{\tilde p}) - {\tilde T} 
s({\tilde T},{\tilde p}) + {\tilde p} \; v({\tilde T},{\tilde p}),
\end{equation}
with ${\tilde T} \equiv T/{\tilde N}$ and  ${\tilde p} \equiv 
p/{\tilde N}$. So, when $\alpha/d >1$, ${\tilde N}$ is a constant, 
and we recover the usual concepts that $G,\;U,\;S$ and $V$ are {\it 
extensive} variables since they scale like $N$, whereas $T$ and $p$ 
are {\it intensive} variables since they are invariant with respect 
to $N$. If, however, $0 \le \alpha/d \le 1$, the 
extensivity-intensivity concepts loose their usual simple meanings. 
Indeed, there are now {\it three}, and {\it not two}, thermodynamical 
categories, namely the {\it energetic} ones ($G,\;F,\;U,\;...$) which 
scale like $N{\tilde N}$, the {\it pseudo-intensive} ones 
($T,\;p,\;H,\;...$) which scale like ${\tilde N}$, and their 
conjugate {\it pseudo-extensive} variables ($S,\;V,\;M,\;...$) which 
scale like $N$ ({\it if and only if} expressed in terms of the 
pseudo-intensive variables). I can unfortunately not prove the 
general validity of the above scheme, but it has been shown to be 
true in {\it all} the models that have been numerically or 
analytically studied \cite{models} (one- and two-dimensional Ising 
and Potts magnets, one-dimensional bond percolation and $XY$ coupled 
rotators, one-, two- and three-dimensional Lennard-Jones-like fluids, 
among others). Also, for the case $\alpha=0$, we recover the 
traditional scalings of Mean Field Approximations, where the 
Hamiltonian is {\it artificially} made extensive by dividing the 
coupling constant by $N$. By so doing, we are obliged to pay a 
conceptually high price, namely to have microscopic couplings which 
depend on $N$! This long standing tradition in magnetism has, for 
very good reasons, never been adopted in astronomy: we are not aware 
of any astronomer dividing the universal gravitational constant $G$ 
by any power of $N$ in order to artificially make extensive a 
Hamiltonian which clearly is not! The thermodynamical scheme that has 
been described above escapes from such unpleasant criticism. I 
sometimes refer to the above discussion of long-ranged systems as 
{\it weak violation} of BG statistical mechanics. Indeed, it becomes 
necessary to conveniently rescale the thermodynamic variables (using 
${\tilde N}$) but the $q=1$ approach remains the adequate one, i.e., 
energy distributions still are of the exponential type.

Let us now address the much more complex $\lim_{t \rightarrow \infty} 
\lim_{N \rightarrow \infty} $ situation. At the present moment, the 
reliable informations that we have are mainly of numerical nature, 
more precisely computational molecular dynamics. A paradigmatic 
system has been intensively addressed during last years, namely a 
chain of coupled inertial planar rotators (ferromagnetic $XY$-like 
interaction) , the interaction decaying as $1/r^{\alpha}$. The system 
being classical, it is thermodynamically extensive for $\alpha>1$ and 
nonextensive for $0 \le \alpha \le 1$. Important informations were 
established by Ruffo, in Florence, and collaborators 
\cite{latorarapisardaruffo}, who focused on the $\alpha=0$ case. If 
we note $E_N$ the total energy associated with $N$ rotators, it can 
be shown that a second order phase transition occurs at the critical 
point $e \equiv E_N/(N{\tilde N})  =3/4 \equiv e_c$. For $e>e_c$ the 
rotators are nearly decoupled and  for $e<e_c$ they tend to 
clusterize. Not surprising, $e_c$ depends on $\alpha$, and vanishes 
above a critical value of $\alpha$ which is larger than unity. For 
$\alpha>1$, nothing very anomalous occurs. But, in the interval $0 
\le \alpha \le 1$, curious phenomena happen on {\it both} sides of 
$e_c(\alpha)$. 

For $e$ {\it above} $e_c$ (say $e=5$), the (conveniently scaled) 
maximum Lyapunov exponent $\lambda_{max}(N)$ goes to zero as 
$N^{-\kappa}$, where $\kappa(\alpha)$ was numerically established, by 
Celia Anteneodo and myself \cite{anteneodotsallis} (see Fig. 3). 
Nothing like this is observed for $\alpha>1$, where the numerical 
evidence strongly points that $\lim_{N \rightarrow \infty} 
\lambda_{max}(N) >0$. This of course suggests that, for a 
thermodynamically large system, mixing is not exponential, but 
possibly a power-law, i.e., weak mixing, as discussed above for the 
one-dimensional maps. The $\kappa(\alpha)$ dependence has been 
recently studied by Andrea Giansanti, in Rome, and collaborators for 
$d=2$ and $3$ as well. If ploted against $\alpha/d$ (see Fig.3), an 
universal curve seems to emerge. This evidence is reinforced by the 
fact that it is possible to analytically prove, for this and other 
models, that $\kappa(0)=1/3$. It is consistently obtained $\kappa = 
0$ for $\alpha/d \ge 1$. 

Let us now focus on the anomalies {\it below} $e_c$. The case 
$e=0.69$ for $\alpha=0$ has been and is being further studied in 
detail. The 
maximal Lyapunov exponent is here positive in the limit $N 
\rightarrow \infty$ (for all $0<e<e_c$ in fact) {\it but}, 
suggestively enough, the usual (BG) canonical ensemble equilibrium is 
preceded by a long, metastable-like, {\it quasi stationary state} 
(QSS), whose duration $\tau$ diverges with $N$! This means that, if 
$N$ is of the order of the Advogadro number, during a time that might 
be longer than the age of the universe, only this curious QSS will be 
observable. So, we better understand it! As can be seen in Fig. 4, 
the average kinetic energy per particle $e_k$ depends on both $N$ and 
time $t$. Only in the $\lim_{N \rightarrow \infty}  \lim_{t 
\rightarrow \infty}$ the BG canonical temperature $T$ is attained. If 
we take the limits the other way around a completely different result 
is obtained, namely $e_k$ approaches a finite value below $T$, and 
$\tau$ diverges like $N^{\rho}$. It becomes possible to speculate 
that $\rho(\alpha)$ decreases from a finite positive value to zero, 
when $\alpha/d$ increases from zero to unity; thereafter, we of 
course expect $\rho = 0$, the system being then a well behaved BG 
one. To make this scenario more robust, consistent evidence is 
available concerning diffusion of the $\alpha=0$ rotators. The 
numerical study of this phenomenon has shown 
\cite{latorarapisardaruffo,antonitorcini} an average second moment of 
the angle increasing like $t^{\beta}$, where $\beta$ depends on both 
$N$ and $t$. For fixed $N$, $\beta>1$ (i.e., superdiffusion) for a 
long time until, at $t = \tau_D$ ($D$ stands for {\it diffusion}), 
makes a crossover to unity (normal diffusion). Analogously to what 
was described above, $\tau_D$ increases like $N^{\sigma}$. Once 
again, it is possible to speculate that $\sigma(\alpha)$ decreases 
from some finite positive value to zero when $\alpha/d$ increases 
from zero to unity. The door remains open to whether there is a 
simple connection between $\sigma$ and $\rho$. All these results 
enable to conjecture that a scenario close to what is shown in Fig. 5 
could indeed occur for some physical systems. Before closing this 
section, let me emphasize that what Fig. 4 strongly suggests is the 
{\it possible} validity of the zeroth principle of thermodynamics 
even {\it out} of a canonical BG scenario! Indeed, for large enough 
$N$, systems can share the same "temperature", thus being at thermal 
equilibrium, though this equilibrium is {\it not } the familiar one. 
This is quite remarkable if we take into account that there is 
nowadays quite strong numerical evidence that the distribution of 
velocities in such systems is far from Gaussian, i.e., the familiar 
Maxwellian distribution of velocities is by all means violated. 

\section{\bf COMPARISON WITH EXPERIMENTAL RESULTS}

During this last decade many types of comparisons have been done 
between the present theory and experimental data. They have different 
epistemological status, and range from simple fittings (with 
sometimes clear, sometimes rather unclear physical insight), up to 
completely closed theories, with no fitting parameters at all. Let me 
review a few among them. 

\subsection{\bf Turbulence} X.-P. Huang and C.F. Driscoll exhibited 
in 1994 \cite{huangdriscoll} some quite interesting nonneutral 
electronic plasma experiments done in a metallic cylinder in the 
presence of an axial magnetic field. They observed a turbulent 
axisymmetric  metaequilibrium state, the electronic density radial 
distribution of which was measured. Its average (over typically 100 
shots) monotonically decreased with the radial distance, disappearing 
at some radius sensibly {\it smaller} than the radius of the 
container 
(i.e., a {\it cut-off} was observed). They also proposed four 
phenomenological theories trying to reproduce the experimentally 
observed 
profile. The {\it Restricted 
Minimum Enstrophy} one provided a quite satisfactory 
first-approximation fitting. Bruce M. Boghosian, in Boston, showed in 
1996 \cite{boghosian} that the Huang 
and Driscoll successful attempt {\it precisely} corresponds to 
the optimization of $S_q$ with $q=1/2$, the necessary cut-off 
naturally emerging from the formalism. Since then, improved versions 
of his calculation \cite{tsallis,celia}, as well as controversial 
arguments \cite{brands}, have been published. The merit of first 
connecting the present formalism to turbulence remains however 
doubtless.

In recent months, Fernando M. Ramos and collaborators, in S\~ao 
Jos\'e dos Campos, Brazil \cite{ramos}, Toshihiko Arimitsu and N. 
Arimitsu, in Tokyo \cite{arimitsu}, as well as Christian Beck, in 
London \cite{beck}, have exhibited very interesting connections with 
fully developed turbulence. Beck's theory is a closed one, with no 
fitting parameters. He obtained a quite impressive agreement with the 
experimental distribution of the differences (at distance $r$) of 
radial velocities, including its well known slight skewness (see Fig. 
6). The entropic parameter is given by $1/(q-1)=1+ \log_2 (r/\eta)$ 
where $\eta$ is the viscosity. The fact that $q$ is not universal 
certainly reminds us the criticality of the two-dimensional $XY$ 
ferromagnet, with which nonextensivity presents in fact various 
analogies.

\subsection{\bf High-energy collisions} Ignacio Bediaga and 
collaborators \cite{becumibeck}, in Rio de Janeiro, have worked out, 
along Rolf Hagedorn's 1965 lines \cite{hagedorn}, a phenomenological 
theory for the distribution of transverse momenta in the hadronic 
jets emerging from electron-positron annihilation after central 
collisions at energies ranging from 14 to 161 Gev. Indeed, since 
early ideas of Fermi, and later of Field and Feynman, a 
thermodynamical equilibrium scenario has been advanced for this 
distribution of transverse momenta. Hagedorn developed a full theory 
based on the BG distribution. The success was only partial. Indeed, a 
central ingredient of the physical picture is that higher collision 
energies {\it do not increase the transverse momenta temperature $T$} 
but increase instead the number of involved bosons that are produced 
(like water boiling at higher flux of energy, where only the rate of 
vapor production is increased, but not the temperature). It is this 
central physical expectance that was not fulfilled by Hagedorn's 
calculation; indeed, to fit the curves associated with increasingly 
high energies, increasingly high values for $T$ become  necessary. 
Furthermore, an {\it inflexion point} clearly emerges in the 
distribution, which by no means can be reproduced by the BG 
exponential distribution. The adequation of only two parameters ($T$ 
and $q$) have enabled Bediaga et al \cite{becumibeck} to fit 
amazingly large sets of experimental data (see Fig. 7). And, in order 
to make the expected picture complete, the central demand of $T$ 
being independent from the collision energy is fulfilled!

Grzegorz Wilk, in Warsaw, and collaborators \cite{wilk}, as well as 
D.B. Walton and J. Rafelski in Tucson \cite{walton}, have provided 
further evidences of the applicability of the present ideas to high 
energy physics, where nonmarkovian processes and long-range 
interactions are commonly accepted hypothesis. Consistent evidence 
has also been provided by D.B. Ion and M.L.D. Ion, in Bucharest, very 
especially in hadronic scattering \cite{ion5}.

\subsection{\bf Solar neutrino problem} Piero Quarati and 
collaborators, in Torino and Cagliari, have been advancing since 1996 
\cite{quarati} an interesting possibility concerning the famous solar 
neutrino problem. Indeed, it is known that calculations within the 
so-called Standard Solar Model (SSM) provide a neutrino flux to be 
detected on the surface of the Earth which is roughly the {\it 
double} of what is actually measured in several laboratories around 
the world. This enigmatic discrepancy intrigues the specialists since 
over 20-30 years ago. At least two, {\it non exclusive}, possible 
causes are under current analysis. One of them concerns the nature of 
neutrinos: they could oscillate in such a way that only part of them 
would be detectable by the present experimental devices. The second 
one addresses the possibility that the SSM could need to be sensibly 
improved in what concerns the production of neutrinos at the solar 
core plasma. It is along this line that Quarati's suggestion 
progresses. Indeed, the neutrino flux is related to the total area of 
the so-called Gamow peak, which is in turn due to the product of the 
thermal equilibrium BG distribution function (which decreases with 
energy) and the penetration factor (which increases with energy). The 
position of the peak is at energies 10 times larger than $kT$, 
therefore only the far {\it tail} of the thermal distribution is 
concerned. Quarati et al argue that very slight departures from $q=1$ 
(of the order of 0.01) are enough to substantially modify (close to 
the desired factor two) the area of the Gamow peak. They also show 
that this degree of departure remains within the experimental 
accuracy of other independent measurements, such as the 
helioseismographic ones. This slight nonextensivity would be related 
to quite plausible nonmarkovian processes and other anomalies.

\subsection{\bf Others} Many other phenomena have been, during recent 
years, temptatively connected to the present nonextensivity. Let me 
briefly mention some of them. The experimental, clearly non Gaussian, 
distribution of velocities in various systems has been shown to be 
very satisfactorily fitted by $q \ne 1$ curves. Such is the case of 
the distribution (observed from the COBE satellite) of peculiar 
velocities of spiral galaxies, shown \cite{marco},  in 1998 by 
Quarati, myself and collaborators, to be very well fitted by $q 
\simeq 0.24$. Such is also the case of the distribution that James 
Glazier, in Notre Dame University, and collaborators \cite{hydra}, 
measured for the horizontal velocities of {\it Hydra Vulgaris} living 
in physiological solution. They were remarkably well fitted using $q 
\simeq 1.5$. Such is finally the case of the computationally 
simulated granular matter involving mesoscopic inelastic collisions. 
Indeed, the distribution function that Y.-H. Taguchi and H. Takayasu 
used to fit their 1995 data \cite{granular} precisely is a typical 
$q>1$ one. 

In 1995 \cite{barreto}, F.C. Sa Barreto, in Belo Horizonte, Brazil, 
E.D. Loh, in Michigan State University, and myself advanced the 
possibility of explaining very tiny departures from the black-body 
radiation Planck's law for fitting the COBE satellite data for the 
cosmic background microwave radiation. Indeed, values of $q$ 
departing from unity by the order of $10^{-5}$ were advanced. This 
possibility has since then be further analyzed by a variety of 
scientists, the conclusion still remaining basically the same. Only 
more precise (perhaps 10 times more precise would be enough) 
experimental data would confirm or reject such a possibility, 
physically motivated by present or ancient subtle effects of 
gravitation, which could even lead ultimately to a modification of 
our understanding of the nature of spacetime (at very small scales, 
it could very well be noncontinuous and even nondifferentiable, in 
contrast with our usual perceptions!).

Molecules like $CO$ and $O_2$ in some hemoproteins can be dissociated 
from their natural positions by light flashes, as many classical 
experiments have shown. Under specific circumstances, these molecules 
then tend to re-associate with a time-dependent rate. The 
experimental rate has been shown, in 1999 by G. Bemski, R.S. Mendes 
and myself \cite{bemski}, to be very well fitted by the solutions of 
a nonextensive-inspired differential equation. The intrinsic 
fractality of such proteins is believed to be the physical motivation 
for such approach. Numbers of citations of scientific publications 
can also be well fitted with $q>1$ curves, consistently with S. 
Denisov's 1997 recovering, within the present formalism, of the 
so-called Zipf-Mandelbrot law for linguistics \cite{denisov}. Oscar 
Sotolongo-Costa, in La Habana, and collaborators \cite{sotolongo}, as 
well as K.K. Gudima, in Caen, France, and collaborators 
\cite{gudima}, have recently invoked the present formalism for 
describing multifragmentation.

Last but not least, it is worthy mentioning that the present 
framework has induced \cite{tsallisstariolo} quite quicker versions 
of the Simulated Annealing methods for global optimization. This 
methodology has found successful applications in theoretical 
chemistry, in particular for studying proteins, their folding and 
related phenomena. Such procedures are currently being used by John 
E. Straub, at the Boston University, and collaborators, by Kleber C. 
Mundim, in Salvador, Brazil, Donald E. Ellis, in Chicago, and 
collaborators, by Yuko Okamoto, in Okazaki,  Ulrich H.E. Hansmann, in 
Houghton, and collaborators, among others \cite{more}.
 
\section{\bf FINAL COMMENTS}

Many issues remain, at the present moment, partially or fully open to 
better understanding concerning the present attempt of adequately 
generalizing Boltzmann-Gibbs statistical mechanics and 
thermodynamics. These include (i) the direct checking of the energy 
distribution which generalizes the BG factor (and consistently the 
connection between $q$ and $(\alpha,d)$ for long-range interacting 
Hamiltonian systems (both finite and large), and similar connections 
for nonconservative ones); (ii) appropriate understanding within the 
renormalization group framework (Alberto Robledo, in Mexico, has 
recently initiated this line \cite{robledo}; see also \cite{RG}); 
(iii) possible firm connections of this formalism with the quantum 
group formalism for generalizing quantum mechanics (the great 
analogies that these two nonextensive formalisms exhibit are since 
long being explored, but not yet deeply understood); (iv) possible 
connections with the L\'evy-like problematic reviewed recently 
\cite{zaslavsky} by George M. Zaslavsky; (v) the strict conditions 
(in terms of the $N \rightarrow \infty$ and $t \rightarrow \infty$ 
limits) under which the zero$^{th}$-principle of thermodynamics 
(i.e., the criterium for thermal equilibrium) holds; (vi) the 
microscopic and mathematical interpretation of the direct and escort 
distributions (which possibly reflect the choice of a (multi)fractal 
description, or of a description in terms of Lebesgue-integrable 
quantities); (vii) the general connection between symmetry and 
entropic form adapted to measure information about systems having 
that particular symmetry; (viii) the rigorous basis for the 
generalization herein described of concepts such as Kolmogorov-Sinai 
entropy, Lyapunov exponents, Pesin equality, escape rates, etc; (ix) 
the possibility of having, for {\it long-range} interacting systems, 
a special value for $q$ which, through the use of appropriately 
rescaled variables, would enable a new kind of extensivity in the 
sense that $S_q(A+B)=S_q(A)+S_q(B)$, where the {\it large} subsystems 
$A$ and $B$ would of course be {\it not independent} (in the sense of 
theory of probabilities) but, on the contrary, strongly coupled; (x) 
the connection between the index $q$ appearing in properties such as 
the sensitivity to the initial conditions, and that appearing in 
properties such as the microscopic distribution of energies (the most 
typical values for the former appear to be {\it smaller} than unity, 
whereas they are {\it larger} than unity for the latter); (xi) the 
direct checking of the type of mixing which occurs in the phase-space 
of many-body long-range interacting large systems; (xii) the deep 
connection with quantum entanglement through its intrinsic 
nonlocality. The last point certainly appears as very promising and 
relevant. An important step forward was very recently done by Abe and 
Rajagopal \cite{aberajagopal}; indeed, they succeeded in recovering, 
through use of the present formalism, the $x=1/3$ Peres (necessary 
and sufficient) condition, which is known to be the {\it strongest} 
one (stronger than Bell inequality) for having local realism (some 
degree of separability of the full density matrix into those 
associated with subsystems $A$ and $B$) for the  bipartite 
spin-$\frac{1}{2}$ system (Werner-Popescu state). The probability of 
this approach being fundamentally correct is enhanced nowadays by 
contributions produced from quite different viewpoints \cite{brukner}.

As we can see, the number of important issues still needing 
enlightening is quite large. This makes the enterprise but more 
stimulating! However, caution is recommended in spite of the sensible 
amount of objective successes of the proposal. Indeed, given the 
enormous impact of the thermodynamical concepts in Physics, one 
should address all these points with circumspection. Nevertheless, an 
important new perspective seems to be already acquired. Nikolai S. 
Krylov showed, half a century ago, that the key concept in the 
foundation of standard statistical mechanics is not ergodicity but 
mixing. Indeed, he showed that the Lyapunov exponents essentially 
control the relaxation times towards BG equilibrium. The entropy 
associated with such systems is vastly known to be the BG logarithmic 
one. What emerges now is an even more fascinating possibility, namely 
the {\it type} of mixing seems to determine the microscopic {\it 
form} of the entropy to be used. If the mixing is of the exponential 
type (strong chaos), then the entropy of course is the BG one (i.e., 
$q=1$). If the mixing is of the power type (weak chaos), then the 
mixing exponent ($1/(1-q)$ apparently) would determine the anomalous 
value of $q$ to be used for the entropy, which would in turn 
determine the thermal equilibrium distribution, as well as all of its 
thermodynamical consequences! The kingdom of the exponentials would 
then be occasionally replaced by the kingdom of the power laws, with 
their relevance in biology, economics and other complex systems!

{\it I thank warm hospitality at the University of North Texas, at 
the Santa Fe Institute and at MIT, where this paper was written. I am 
especially grateful to M. Gell-Mann, who made my visit to the Santa 
Fe Institute possible, stimulating and pleasant. My sincere thanks go 
also to L. E. Reichl whose editorial interest and dedication greatly 
improved the entire Proceedings volume, and very especially the 
present manuscript. Work in my group at the Centro Brasileiro de 
Pesquisas Fisicas, Rio de Janeiro, is supported in part by 
PRONEX/FINEP, CNPq and FAPERJ (Brazilian Agencies).}\\

{\bf APPENDIX}\\

Scene at the restaurant\\

{\it AU LABYRINTHE DES ENTROPIES} \\

{\it -- Would you have some fresh entropies today, for me and my 
friends?

-- Absolutely Sir ! 
We have them extensive or not, with definite concavity or not, 
nonnegative defined or otherwise, quantum, classical, relative, cross 
or mutual, included in several others with a small supplement, 
composable or not, expansible or not, totally optimized or a little 
rare, even completely out of equilibrium...  single-trajectory-based 
or ensemble-based...

You can have them at the good old Boltzmann magnificent style, 
dor\'ee \`a la Gibbs, very subtle, or von Neumann...,  with pepper 
\`a la Jaynes, or the popular Shannon, oh, my God, I was forgetting 
the esoteric, superb, macroscopic Clausius, the surprising Fisher, 
the refined Kolmogorov-Sinai, \`a la Kullback-Leibler for comparison, 
Renyi with multifractal dressing, with cybernetic sauce \`a la Harvda 
and Charvat, Vajda and Daroczy, or even the all-taste Sharma and 
Mittal...

In 1988 we started serving them with Brazilian touch, if you wish to 
try, it leaves a tropical arri\`ere-gout in your mouth! And since 
then, our chefs have introduced not less than ten new recipes... 
Curado, with exponentials, Anteneodo with tango flavor, Plastino's, 
excellent as family dish, Landsberg, Papa, Johal with curry, Borges 
and Roditi, Rajagopal and Abe... 

They are all delicious ! How many do I serve you today?}\\

{\bf FIGURES}\\

{\bf Fig. 1 -} Equilibrium-like probability versus energy for typical 
values of $q$, namely, from top to bottom at low energies, 
$q=0,\;1/4,\;1/2,\;2/3,\;1,\;3,\; \infty$ (the latter collapses onto 
the ordinate at the origin, and vanishes for all positive energies). 
For $q=1$ we have Boltzmann exponential factor; for $q>1$ we have a 
power-law tail; for $q<1$ there is a cutoff above which the 
probability vanishes. \\

{\bf Fig. 2 -} Time evolution of $S_q$: (a) $a=2$ ($q^*=1$); (b) 
$a=a_c$ ($q^*=0.2445$; $R$ measures the degree of nonlinearity of the 
curves in the relevant intermediate region).\\

{\bf Fig. 3 -} Coupled planar rotators: $\kappa$ versus $\alpha/d$ 
($d=1$: \cite{anteneodotsallis}; $d=2,3$: \cite{giansanti}). The 
solid line is a guide to the eye.\\

{\bf Fig. 4 -} Coupled planar rotators ($\alpha=0$, $d=1$): Time 
evolution of $T$. From Latora and Rapisarda \cite{rapisardadenton}).\\

{\bf Fig. 5 -} Conjectural time dependence of the probability $p$ of 
having an energy $E$ for classical systems (${\tilde N} \equiv 
N^*+1$).\\

{\bf Fig. 6 -} Distribution of radial velocity differences in fully 
developed turbulence (from top to bottom: $r/\eta = 3.3,23.6,100$). 
From Beck \cite{beck}.\\

{\bf Fig. 7 -} Distribution of hadronic transverse momenta: (a) 
cross-section (dotted line: Hagedorn's theory; solid lines: $q \ne 1$ 
theory); (b) Fitting parameters $q$ and $T_0$. From Bediaga et al 
\cite{becumibeck}.\\

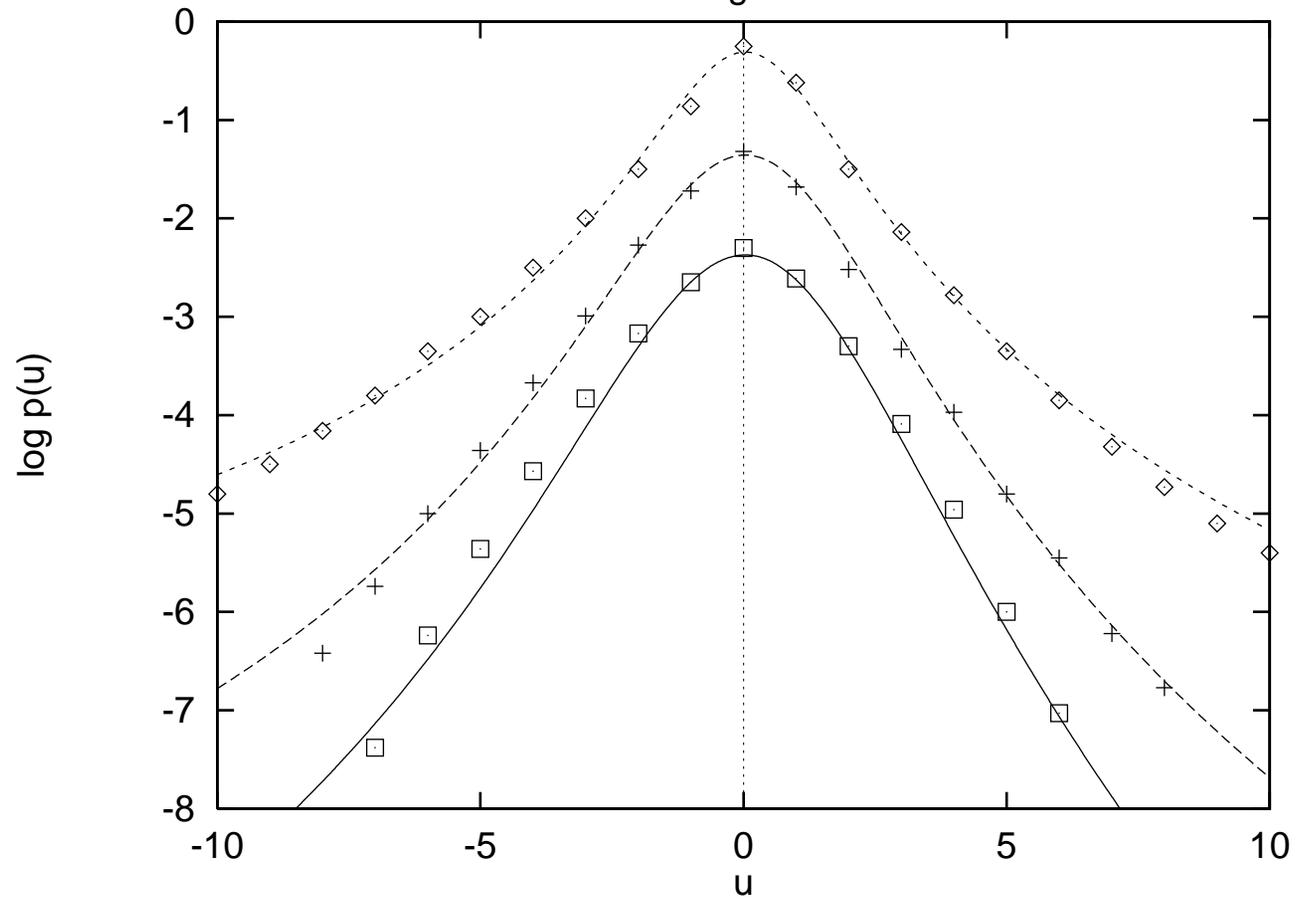
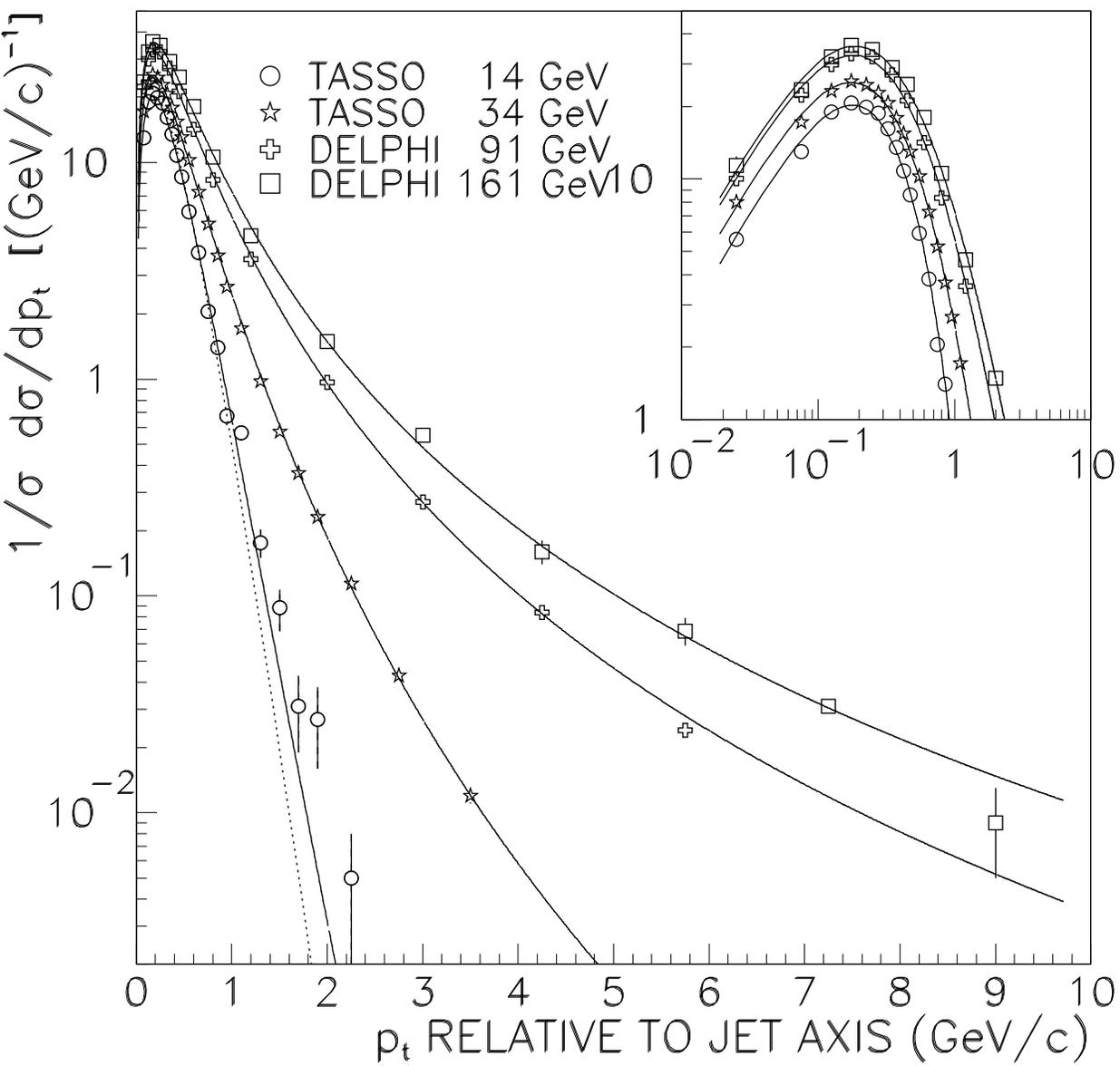
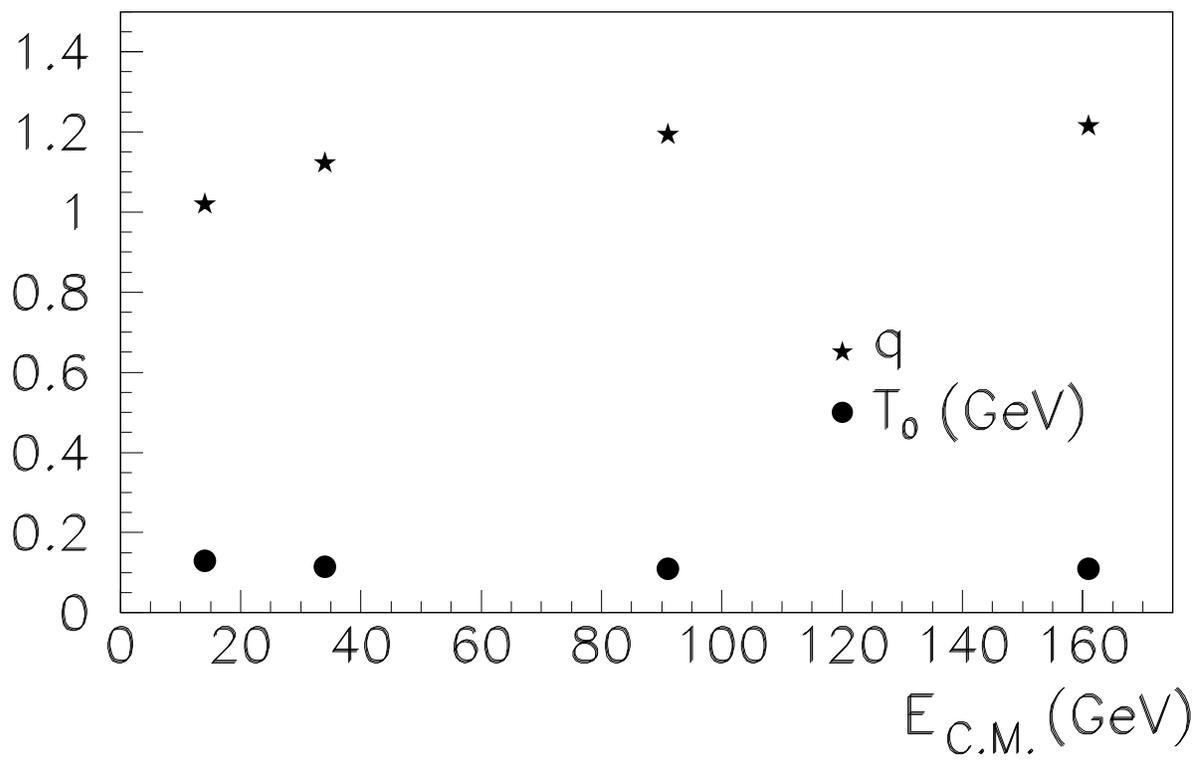
\end{document}